# Architecture Without an Architect? Global Governance of Artificial Intelligence in a Divided World

Review Essay

MATTHIJS M. MAAS, *Architectures of Global AI Governance: From Technological Change to Human Choice* (Oxford: Oxford University Press, 2025)

*Simon Chesterman, National University of Singapore*

Artificial intelligence presents an unusually difficult problem for global governance. The technology develops rapidly, crosses borders easily, and is shaped by actors whose resources and capabilities may rival those of states. Yet international responses remain fragmented, unevenly representative, and overwhelmingly non-binding. The challenge is therefore not simply to identify appropriate rules or institutions, but to understand who has the capacity and incentive to create, enforce, and adapt them.

This review essay examines these questions through Matthijs Maas's *Architectures of Global AI Governance*. Maas offers an ambitious framework for thinking about AI governance through the lenses of sociotechnical change, governance disruption, and regime complexity. His account usefully resists both technological determinism and the search for a single institutional blueprint, emphasizing instead the possibilities of a fragmented and evolving governance architecture.

The essay argues, however, that institutional design cannot be separated from the distribution of power. Maas frequently invokes what "we" should do about AI, but that collective subject obscures important differences among states, international institutions, and technology companies. States retain formidable powers over markets, infrastructure, strategic inputs, and firms themselves. At the same time, many consequential decisions about frontier AI — what is built, how quickly, with what safeguards, and when it is released — are concentrated within a small number of private companies. The central problem of global AI governance may therefore be less architecture without an architect than an emerging architecture shaped by multiple actors possessing different forms of power, divergent incentives, and no common set of plans.

These are boom times for AI governance. An ever-expanding number of companies and countries have adopted hundreds of guides, frameworks, and principles that purport to moderate the risks associated with a fast-moving technology that could improve the world or destroy it — with a surprising number of technopreneurs holding both Panglossian and catastrophist positions simultaneously. The patchwork of national initiatives is now complemented by a growing canvas of regional efforts, notably including the various European architectures, as well as initiatives in Africa, the Americas, and Asia. Selective groups like the G7, the G20, the OECD, and the Global Partnership on AI have also adopted foundational documents. Private sector bodies such as the World Economic Forum and standards organizations have been active, as have civil society actors. Even the Pope has entered the fray, placing AI high on his own agenda.[1]

Yet the most important documents have hardly been 'global'. In 2024, the United Nations estimated that only seven states were party to the seven most important such initiatives — the wealthy, industrialized members of the G7 — while fully a hundred other states were party to none of them.[2] To the armchair regulator, the need for a global response is self-evident. The technology pays no heed to boundaries; to the extent that it does, companies are apt to engage in regulatory arbitrage to seek more permissive jurisdictions.

This has led to a proliferation of academic proposals for regimes of various kinds in journals and books. There is no shortage of models to draw upon when seeking to establish some form of global architecture. Popular candidates include bodies such as the International Atomic Energy Agency (IAEA),[3] the European Organization for Nuclear Research (CERN),[4] and the

[1] Pope Leo XIV, *Encyclical Letter Magnifica Humanitas of His Holiness Pope Leo XIV on Safeguarding the Human Person in the Time of Artificial Intelligence* (The Holy See, 15 May 2026). See generally Simon Chesterman et al, 'The Evolution of AI Governance' (2024) 57(9) IEEE Computer 80.

[2] UN AI Advisory Body, *Governing AI for Humanity: Final Report* (United Nations, September 2024) 9. Disclosure: the author served as principal researcher for the UN AI Advisory Body at the time the report was drafted.

[3] Simon Chesterman, 'Weapons of Mass Disruption: Artificial Intelligence and International Law' (2021) 10 Cambridge International Law Journal 181.

[4] Anna-Lena Rüland, '"We Need a CERN for AI": Organized Scientific Interests and Agenda-Setting in European Science, Technology, and Innovation Policy' (2025) Minerva.

Intergovernmental Panel on Climate Change (IPCC).[5] Yet when the UN actually came to discuss the matter concretely at the Summit of the Future in 2024, there was no appetite and even less money to do anything architecturally ambitious.[6]

Fast-moving technology puts regulators in an unenviable position. Act too swiftly and they constrain innovation or drive it to more permissive jurisdictions. Act too slowly and they expose their citizenry to risk. Recent history is littered with examples of both, from the US falling behind in stem-cell research after Christian lobbyists secured a moratorium in 2001[7] to the global struggle to address social media harms following two decades of inaction.[8] Even when they do act, the response is often ‘reactive, fragmented, and grounded in shallow analogies to past technologies or legacy institutions’.[9] For their part, technology companies themselves have increasingly accepted the need for some form of regulation in theory — even as they oppose any specific proposals in practice.[10]

Into this space, Matthijs M. Maas offers a weighty new tome: *Architectures of Global AI Governance*. He argues that we must reject the ‘epistemic and political pessimism’ (594) that often paralyses debates over AI governance: assuming that the technology’s future development is too speculative or too inevitable for us to make meaningful decisions today. Maas, now Senior Research Fellow at the Institute for Law & AI in Boston, argues that global governance of AI can and should ‘do better’ (2) by drawing on the concepts of sociotechnical change, governance disruption, and regime complexity.

Writing on any dynamic field risks falling into the relentless presentism that characterizes a lot of journalism about AI, with breathtaking advances and dangers announced almost on a weekly basis. Maas takes a longer horizon. His book builds on a doctoral thesis at the University of Copenhagen Faculty of Law, pursued under the supervision of Hin-Yan Liu and Kristian Lauta and completed in 2021. The work has been updated and expanded to embrace the mass adoption of AI after the arrival of ChatGPT in late 2022, as well as the excitement over agentic AI that reached fever pitch in 2025.

---

5 Joseph Bak-Coleman et al, 'Create an IPCC-like Body to Harness Benefits and Combat Harms of Digital Tech' (2023) 617(7961) Nature 462.

6 *Global Digital Compact* (United Nations, September 2024).

7 Varnee Murugan, 'Embryonic Stem Cell Research: A Decade of Debate from Bush to Obama' (2009) 82 Yale Journal of Biology and Medicine 101.

8 Robert Gorwa, *The Politics of Platform Regulation: How Governments Shape Online Content Moderation* (New York: Oxford University Press, 2024).

9 Matthijs M. Maas, *Architectures of Global AI Governance: From Technological Change to Human Choice* (Oxford: Oxford University Press, 2025) 2.

10 Jonas Tallberg, Magnus Lundgren, and Johannes Geith, 'AI Regulation in the European Union: Examining Non-State Actor Preferences' (2024) 26(2) Business and Politics 218; Simon Chesterman, 'Silicon Sovereigns: Artificial Intelligence, International Law, and the Tech-Industrial Complex' (2026) 120(1) American Journal of International Law 44.

The finished work now runs to more than 600 pages, its 3,000-plus footnotes on occasion threatening to overwhelm the text.[11] Maas draws on an eclectic range of disciplines, from law and political science to sociology and the history of technology. In words that will resonate with any former doctoral student, he rails against ‘space constraints’ (91), ‘insufficient space’ (349), the need for ‘future work’ (581), and bemoans on the penultimate page of the conclusion — the 600th page of his book — that he was unable to ‘explore topics as deeply as would have been ideal’ (600).

Another trope common to theses is that he is, on occasion, overly modest in his goals. A deflecting footnote stresses that his aim is merely to ‘provide a set of exploratory perspectives and tools’ (38 n222); over the course of the volume, he makes good on his promise not to ‘offer up simple, definitive, policy blueprints’ (600). That might lead some readers to question their investment in acquiring the book, though its open access status means that they would likely fail to obtain any refund.

His aim, instead, is to contribute ‘to both an improved analytical understanding of AI governance issues and to better-informed international policy and institutional development’ (283). Among the strengths of the book is the manner in which it reflects not only on how AI poses a challenge for global governance structures, but how it is changing those structures also. Through this approach, he argues, ‘we can adapt our global governance proposals for advanced AI in a manner that is more robust to ongoing and future changes in AI technology, in law, and in international politics’ (3).

That ‘we’ is doing some heavy lifting, however. From an analytical shorthand, the term is soon deployed also as a rhetorical call to action: ‘It is important that *we* get AI governance right, and that *we* do so soon — as *we* may not have many future chances to do so’ (21, emphases added). A quick search finds the word ‘we’ more than 400 times in the text. Why ‘we’ need global governance of AI, however, is not the same as explaining why those who actually wield power in this space — states, but increasingly technology companies themselves — might actually create it. Nor is power distributed uniformly between them. States retain formidable coercive and regulatory capacities, while frontier AI companies increasingly control the technical decisions, infrastructure, and expertise on which effective governance depends. The problem with ‘we’ is therefore not that nobody has power, but that those who do have it possess different kinds of power and may have little incentive to exercise them towards a common end. Indeed, it is striking that, in a 600-page book about the architectures of global AI governance, there is not a single mention of an architect.

[11] It is meticulously organized: one section is numbered 2.2.2.2.2, though any pleasure at the symmetry of its subheading is effaced by the description of an ‘artificial intelligence enabled kill web’ in the context of military applications of AI (129).

# 1 Lenses

The book is structured in three parts. It begins with a survey of the challenges posed by AI, the need for a global response, and how initiatives to date have fallen short. Part II seeks to draw lessons from sociotechnical change, governance disruption, and regime complexity to better understand the current moment and the limitations of existing approaches. The final part seeks to set out ‘frameworks for choice’ in order to help shape an emerging AI governance regime that, at its best, might be effective, resilient, and coherent.

The lens of sociotechnical change offers potential guidance on allocating scarce attention of regulators to the most socially disruptive technologies (371). It also suggests past examples that demonstrate the pros and cons of targeting that attention upstream (for example, to model developers) or downstream (to deployers, users, and so on). Drawing on Collingridge, it can also help address the question of *when* to act — while acknowledging the dilemma that early action is easy but uncertain, whereas delay allows greater clarity about risks but increases the costs of addressing them.[12]

When turning to the disruptive impact of AI on governance, Maas sensibly does not overplay his hand. Indeed, unlike technologies that truly opened up *terra incognita* — space travel, the early days of cyberspace — AI is accelerating and transforming, for the most part, activities that map onto traditional jurisdictions (392). That is not to say that AI does not pose regulatory challenges. Its speed, autonomy, and increasing opacity pose many practical and some conceptual difficulties for existing norms. There are gaps, to be sure — maritime law long defined a ship as a ‘manned’ [*sic*] vessel;[13] traffic law in many states criminalizes leaving an automobile unattended.[14] These must be amended if autonomous vessels and vehicles are to ply our waterways and our roads. But it would be too much to say that AI requires a complete overhaul of our conceptions of governance and regulation.[15] Indeed, early efforts to regulate the evolving world of cyber activities also offer examples of over-extension. The 1984 US Computer Fraud and Abuse Act (CFAA), on a literal reading, threatened jail terms for innocuous activities such as sharing a password with one's spouse or accessing non-work websites from a company computer.[16]

[12] David Collingridge, *The Social Control of Technology* (London: Frances Pinter, 1980).

[13] Robert Veal and Michael Tsimplis, 'The Integration of Unmanned Ships into the *Lex Maritima*' (2017) [2017] Lloyd's Maritime and Commercial Law Quarterly 303, 308–14.

[14] See, e.g., Road Vehicles (Construction and Use) Regulations 1986 (UK), reg 107.

The third lens Maas employs is that of regime complexity, which is the interaction of multiple global governance mechanisms with overlapping mandates and impacts.[17] This has clear applications to the emerging and fragmented field of AI governance at the global level in particular. It also points to the manner in which polycentric governance allows for a measure of experimentation and inter-institutional learning,[18] as well as reflecting the pluralism of a diverse and multipolar world.[19] Interesting, but a very glass-half full analysis of international efforts at a time when the legitimacy of international law itself may be at its lowest ebb.[20]

In the final substantive chapter, Maas promises to draw the preceding excursions together into an 'analytical framework or toolkit' (497). With regard to advanced AI, Maas is sceptical of efforts to ban AI *techniques*, which are either dual-use or do not map neatly onto identifiable harms (530). Red lines or bans have historically been more effective but raise in turn the problem of how to specify the nature of the harm being avoided — though some, such as the use of military AI in nuclear command and control, appear to be gaining traction (529). He proposes to focus on AI *capabilities*. In the military domain, this yields some interesting discussion of how the defence community conceives of AI and its impact on security, the extent to which various emerging applications are 'regulation-tolerant' or '-resistant' (504), as well as comparisons with past efforts at arms control. Those efforts are said to offer reasons for both optimism and pessimism (518), though his account of how epistemic communities reframed emerging innovations as establishing global norms that later forms the basis for international cooperation and institutionalisation requires tolerance for movement through 'smaller, informal, and like-minded groups' being empowered to settle the norms, even if they fragment the regime complex (524).

[15] Indeed, there are ways in which AI could enhance governance of other global challenges. Maas surveys potential use cases in monitoring everything from arms control and human rights to environmental law (441-453).

[16] Tim Wu, 'Fixing the Worst Law in Technology', *New Yorker* (18 March 2013).

[17] See generally Kal Raustiala and David G. Victor, 'The Regime Complex for Plant Genetic Resources' (2004) 58(2) International Organization 277; Amandine Orsini, Jean-Frédéric Morin, and Oran Young, 'Regime Complexes: A Buzz, a Boom, or a Boost for Global Governance?' (2013) 19(1) Global Governance 27. As originally formulated, a lack of obvious hierarchy between them rendered their interactions more complex, though this was later debated as to its necessity at the level of definition.

[18] Karen J. Alter and Sophie Meunier, 'The Politics of International Regime Complexity' (2009) 7(1) Perspectives on Politics 13; Sebastian Oberthuer, 'Interplay Management: Enhancing Environmental Policy Integration Among International Institutions' (2009) 9(4) International Environmental Agreements: Politics, Law and Economics 371.

[19] Amitav Acharya, 'The Future of Global Governance: Fragmentation May Be Inevitable and Creative' (2016) 22(4) Global Governance 453.

[20] Oona A. Hathaway and Scott J. Shapiro, 'Might Unmakes Right' (2025) 104(4) Foreign Affairs 80.

Again, the 'we' is implied. Though it is also a problem that such developments take time, which the rest of 'us' might not have.

# 2 Action and Reaction

The harsh truth is that major change in the institutions of global order requires crisis. We would not have had the League of Nations without the First World War; the United Nations would not have existed but for the Second. The IAEA emerged from the shadows of Hiroshima and Nagasaki,[21] much as the modern taboo against chemical weapons owes much to the trenches of Ypres and Passchendaele.[22] Swift action is sometimes possible — the Montreal Protocol to address ozone depletion is a case in point, arguably the most successful environmental agreement in history.[23] Yet the larger climate crisis offers a stark warning: it is a slow-moving crisis with fairly clear scientific consensus about the problem, and yet action remains difficult. AI offers a fast-moving challenge with no consensus and even less willingness to act in a coordinated manner.

'We' are also reactive in inconsistent ways. Despite widespread fears of lethal autonomous weapon systems (LAWS) being normalized, their use by Ukraine has been quietly tolerated by many governments because of sympathy for its cause against Russian aggression.[24] Some international lawyers similarly tied themselves in knots to support NATO's 1999 intervention in support of autonomy (and later independence) for Kosovo, arguing that it was 'illegal but legitimate' — arguments that were subsequently cited by Russia in its subsequent purported annexation of Crimea and indeed in support of its 2022 invasion.[25]

---

[21] Robert L. Brown, *Nuclear Authority: The IAEA and the Absolute Weapon* (Washington, DC: Georgetown University Press, 2015).

[22] Richard M. Price, *The Chemical Weapons Taboo* (Ithaca: Cornell University Press, 1997).

[23] Edward A. Parson, *Protecting the Ozone Layer: Science and Strategy* (New York: Oxford University Press, 2003).

[24] Ingvild Bode, 'Emergent Normativity: Communities of Practice, Technology, and Lethal Autonomous Weapon Systems' (2024) 4(1) Global Studies Quarterly; Michal Onderco, 'Navigating the AI Frontier: Insights from the Ukraine Conflict for NATO's Governance Role in Military AI' (2025) 48(3) Journal of Strategic Studies 602.

Similar reactivity is evident in response to the threats and opportunities posed by AI. Witness the whiplash from the AI Safety Summit at Bletchley Park in 2023, launching Safety Institutes in multiple continents. Two years later, the most prominent of these — those in Britain and the United States — had been renamed and the 2025 meeting abandoned AI Safety in favour of 'AI Action', described by many attendees as less of a forum on managing AI than a trade show promoting it.

At times, Maas's wariness of offering simplistic solutions makes it hard to see the problem he is trying to solve. The global deficit in AI governance is the absence of red lines, in particular with regard to uncontrollable and uncontainable AI, which would threaten to escape human capacities to mitigate.[26] To this it could be added that minimal transparency rules should ensure that AI systems remain attributable for their actions back to human, corporate, or state actors. These are the structural challenges posed by AI to existing legal orders. Norms against bias, requirements for safety and the like can be upheld merely by ensuring that existing human rights norms and product liability laws are made to apply. More general calls for accountability are, in essence, demands that AI not be allowed to exist in a normative vacuum. There are, to be sure, larger questions about how the economic disruption of AI will be navigated, how its benefits might be distributed, how the consequences of mass disinformation will be navigated.

But Maas is particularly interested in the global part of global governance. He frames it as a collective action problem, where cooperation can reduce regulatory uncertainty and arbitrage, comparing it to governance of the internet and nuclear non-proliferation (162-165). He also acknowledges the potential for treating AI governance as a global public good, though this is framed primarily in terms of establishing standards for behaviour and constraining bad actors (165-168).

The focus is on three broad areas. 'Conventional' denotes the most widespread forms of AI in civilian domains, typically the focus of most near-term governance discussions. 'Military' embraces the international security dimensions of AI. 'Advanced' points to the wider potential

[25] Independent International Commission on Kosovo, *The Kosovo Report* (Oxford: Oxford University Press, 2000); Marko Milanovic, 'Crimea, Kosovo, Hobgoblins and Hypocrisy', *EJIL: Talk* (20 March 2014).

[26] Nick Bostrom, *Superintelligence: Paths, Dangers, Strategies* (Oxford: Oxford University Press, 2014). Concerns about cybersecurity in particular, after the Claude Mythos Preview release, were showing signs of changing this calculus in mid-2026. See, e.g., Jen Easterly, 'Trump Is Finally Facing Reality on A.I.', *New York Times* (4 June 2026)

for misuse, mistake, and systemic risks of the kind tracked by the various AI safety initiatives (120).[27]

Much of the discussion of these three areas takes place in discrete epistemic communities. The military track saw the first efforts at global regulation, beginning around 2009 with the creation of the International Committee for Robot Arms Control, building on a growing academic literature in the field[28] and later being picked up by mainstream civil society organizations.[29] A UN Special Rapporteur called for a moratorium on robotic and AI weapons in 2013,[30] helping to move the topic onto the agenda of the states parties to the Convention on Certain Conventional Weapons (CCW) in 2014, and in turn over to a Group of Government Experts (GGE) that has met annually since 2017. Progress has been halting, at best, held up in particular by definitional debates over what constitutes a 'lethal autonomous weapon system' and 'meaningful human control'. High profile statements by the UN Secretary-General[31] and a 2023 General Assembly resolution[32] drew attention to the topic. Nonetheless, willingness to move even at the level of non-binding commitments tended to vary inversely with capacity to deploy such systems. Maas delicately notes that current developments stand at a 'crossroads' (191).

Governance of conventional systems entered mainstream discussion around 2016, in particular after the Cambridge Analytica scandal revealed two years later that the potential impact of algorithms went beyond eccentric recommendations on Amazon and occasional identity theft.[33] At the UN level, discussion moved into wider forums with the International Telecommunications Union organizing the first AI for Good summit in 2017, even as the alphabet soup of UN agencies, funds, and programmes began to put AI on the agenda. With regard to governance in particular, UNESCO took a lead role and adopted non-binding recommendations on the ethics of AI in 2021.[34] There have been various efforts at treaties,[35] but the majority of hard regulation has been state-based, most notably the EU AI Act, proposed in 2021[36] and formally adopted three years later.[37]

[27] See, e.g., *International AI Safety Report 2026* (Department for Science, Innovation and Technology and AI Safety Institute, February 2026).

[28] See, e.g., P. W. Singer, *Wired for War: The Robotics Revolution and Conflict in the Twenty-First Century* (New York: Penguin, 2009); Ronald Arkin, *Governing Lethal Behavior in Autonomous Robots* (Boca Raton: CRC, 2009).

[29] *Losing Humanity: The Case Against Killer Robots* (Human Rights Watch, 2012).

[30] Report of the Special Rapporteur on Extrajudicial, Summary, or Arbitrary Executions, Christof Heyns, UN Doc A/HRC/23/47 (2013).

[31] *Our Common Agenda Policy Brief 9: A New Agenda for Peace* (United Nations, July 2023) 27.

[32] GA Res 78/241 (2023).

[33] Editorial, 'Cambridge Analytica Controversy Must Spur Researchers to Update Data Ethics' (2018) 555 Nature 559.

More serious consideration of how to address advanced models only took off around 2022 with the advent of ChatGPT, at which point publics and their politicians began to pay attention as AI moved from a niche topic to become one of the most talked about technologies in history. Hurried efforts to incorporate generative AI into the EU's AI Act demonstrated the challenge of deliberative regulation of accelerating technology. At the global level, the UN Secretary-General created a new High-Level Advisory Body on AI, which issued reports in 2023[38] and 2024,[39] including recommendations adopted into the Global Digital Compact at the Summit of the Future later that year.[40] Among other things, this established Global Policy Dialogues and a new International Scientific Panel that convened for the first time in July 2026.

One of the strengths of the book is Maas's effort to bridge those diverse communities. Shortly after its publication, a public spat between the US Department of Defense and Anthropic over the use of its general purpose Claude model in military applications — in particular, CEO Dario Amodei's request that the model not be used for lethal decisions or mass domestic surveillance — suggested the manner in which those community distinctions might become further blurred.[41]

For all the storm and fury, Maas himself acknowledges that the vast majority of normative movement at the global level is non-binding. This is not necessarily a bad thing. Human rights norms emerged over the course of the twentieth century in significant part because of a conscious effort to start with commitments that were wider in participation than they were deep in terms of obligations.[42] He also notes that much of the international discussion is 'under-inclusive' (223), which is a discreet way of acknowledging that the vast majority of states are not directly involved. Yet the Global South gets only one substantive paragraph (224); even China only merits a few pages (211-213).[43]

[34] *Recommendation on the Ethics of Artificial Intelligence* (UNESCO, 23 November 2021).

[35] See, e.g., Council of Europe Framework Convention on Artificial Intelligence, done at Vilnius, 5 September 2024.

[36] *Proposal for a Regulation of the European Parliament and of the Council Laying Down Harmonised Rules on Artificial Intelligence (Artificial Intelligence Act) and Amending Certain Union Legislative Acts* (European Commission, COM/2021/206 final, 2021).

[37] Regulation (EU) 2024/1689 of the European Parliament and of the Council of 13 June 2024 Laying Down Harmonised Rules on Artificial Intelligence (Artificial Intelligence Act) 2024 (EU).

[38] UN AI Advisory Body, *Interim Report: Governing AI for Humanity* (United Nations, 2023).

[39] UN AI Advisory Body, *Governing AI for Humanity: Final Report* (n #).

[40] *Global Digital Compact* (n #).

[41] Jessica Dorsey et al, 'The Pentagon/Anthropic Clash Over Military AI Guardrails', *Opinio Juris* (26 February 2026).

Arguments for greater involvement include considerations of equity and the need to understand the diverse manner in which technology may affect diverse communities. The strongest argument, perhaps, is that for any regime to be effective it will need to be global — analogous to other collective action problems like climate and pandemic preparedness, or guarding against bad actors that can shift or mask their location, analogous to terrorism and cybersecurity. Those arguments hold significant sway in bodies like the UN; when it was India's turn to hold the next iteration of the AI safety summit, it emphasized impact and distribution of benefits among nations.

Yet the bigger gap may be the assumption that AI governance is something that can be handled by 'we, the peoples' at all.

# 3 The Trillion Dollar Question

The structural barrier to AI governance is not that states have ceased to matter, but that their authority no longer maps neatly onto the sources of power that matter most. States can regulate markets, control strategic inputs, and impose costs. But many of the most consequential decisions about frontier AI — what is built, how quickly, with what safeguards, and when it is released — are made inside private firms. Tellingly, at its Summit of the Future in 2024, member states of the United Nations adopted a Global Digital Compact. Those states committed to various actions by 2030, including ensuring that legislation relevant to digital technologies is ‘compliant with obligations under international law, including international human rights law’.[44] In the same document, those states merely ‘call[ed] on’ technology companies and developers to ‘respect international human rights and principles’.[45] Such is the challenge of implementing norms adopted by states to the private sector.[46]

[42] Jochen von Bernstorff, 'The Changing Fortunes of the Universal Declaration of Human Rights: Genesis and Symbolic Dimensions of the Turn to Rights in International Law' (2008) 19(5) European Journal of International Law 903; Kenneth W. Abbott and Duncan Snidal, 'Hard and Soft Law in International Governance' (2000) 54(3) International Organization 421.

[43] I concede that this review may be accused of the same omissions. But see further Nanjala Nyabola, *Digital Democracy, Analogue Politics: How the Internet Era is Transforming Politics in Kenya* (London: Zed Books, 2018); Payal Arora, *The Next Billion Users: Digital Life Beyond the West* (Cambridge, MA: Harvard University Press, 2019); Damian Okaibedi Eke, Kutoma Wakunuma, and Simisola Akintoye (eds), *Responsible AI in Africa: Challenges and Opportunities* (Springer 2023); Tshilidzi Marwala, *The Balancing Problem in the Governance of Artificial Intelligence* (Singapore: Springer, 2024); Gorden Moyo, *An Anatomy of AI Surveillance in Africa: The Contradictions of Digital Technologies* (Cham: Springer, 2025); Luca Belli and Walter Britto Gaspar (eds), *AI from the Global Majority: Official Outcome Report of the UN IGF Data and Artificial Intelligence Governance Coalition* (Springer 2026).

Maas does acknowledge that ‘there is a strong private sector dynamic’ (564) in debates over AI governance, though this is seen primarily through the lens of participation in the various state-based global institutions, whether they be centralized or fragmented. It is striking that in his discussion of disruption to governance, he barely touches on the most important limitation on the ability of existing institutions to manage the risks associated with AI: the shift in power from public to private hands. Though Maas might respond that his focus is on the technology, rather than the people and the corporations developing it, fleeting discussion of regulatory capture (567-568) does not adequately address the greatest barrier to national and international coordinated action: the fact that the most important chokepoints in frontier AI — compute, cloud, model development, and deployment — are concentrated in a small number of firms, embedded in a small number of jurisdictions, and controlled by a handful of men.[47] At a time when these ‘silicon sovereigns’ are not merely developing world-changing technologies, but doing so in a manner that has made vast proportions of the economy dependent on their success, ‘strong … dynamic’ is something of an understatement.

Some of those chokepoints do reveal the continuing power of at least some states. US export controls on advanced semiconductors have demonstrated Washington’s ability to constrain access to the compute required to develop frontier systems. The Biden Administration went further in January 2025, adopting controls on the weights of certain advanced closed-weight models. Yet that measure also illustrates the contingency of state power: the succeeding administration declined to enforce the AI Diffusion Rule and moved to rescind it, while retaining and adapting controls on advanced chips. At the time of writing, cybersecurity concerns had prompted the Trump Administration to provide for pre-release government scrutiny of frontier models, though this 'regulation' was voluntary and its benchmarking process classified.

[44] *Global Digital Compact* (n #) para 23(a).

[45] Ibid para 25(a).

[46] See generally John Gerard Ruggie, Guiding Principles on Business and Human Rights: Implementing the United Nations 'Protect, Respect and Remedy' Framework, UN Doc A/HRC/17/31 (2011).

[47] See, e.g., Tejas N. Narechania and Ganesh Sitaraman, 'An Antimonopoly Approach to Governing Artificial Intelligence' (2024) 43(1) Yale Law & Policy Review 95; Karen Hao, *Empire of AI: Dreams and Nightmares in Sam Altman's OpenAI* (New York: Penguin, 2025); Anton Korinek and Jai Vipra, 'Concentrating Intelligence: Scaling and Market Structure in Artificial Intelligence' (2025) 40(121) Economic Policy 225; James Muldoon, Ana Valdivia, and Adam Badger, 'The Politics of Artificial Intelligence Supply Chains' (2026) 41(2) AI & Society 1175.

Towards the end of his book is a section entitled 'strategies', though Maas is once more at pains to stress that he is not offering a single roadmap. Indeed, he abjures even discussing 'substantive values or goals for AI regimes' as these may change over time (581). The admirable intent is to provide a future-proof toolbox of governance options for the future, along with a means of evaluating the efficacy, resilience, and coherence of any such regimes.

Maas's achievement is to show that global AI governance is not a blank page. It is already an emerging architecture: fragmented, overlapping, and shaped by the interaction of technology, law, and institutional choice. The harder question is whether anyone is capable of building upon it. States still speak the language of sovereignty and public authority, but the most consequential decisions about frontier AI are increasingly made by private firms whose resources, incentives, and time horizons sit uneasily with the slow rituals of international law — and indeed of domestic law also. The result is not simply architecture without an architect. It may be architecture in which neither the builders nor anyone else is holding the plans.